\documentclass[twoside]{article}

\usepackage{amssymb}
\usepackage[reqno]{amsmath}
\usepackage{graphicx}
\usepackage{subfigure}

\input epsf     

\catcode`\@=11
\long\def\@makefntext#1{
\protect\noindent \hbox to 3.2pt {\hskip-.9pt  
$^{{\eightrm\@thefnmark}}$\hfil}#1\hfill}               

\def\@makefnmark{\hbox to 0pt{$^{\@thefnmark}$\hss}}    
        
\def\ps@myheadings{\let\@mkboth\@gobbletwo
\def\@oddhead{\hbox{}
\rightmark\hfil\eightrm\thepage}   
\def\@oddfoot{}\def\@evenhead{\eightrm\thepage\hfil
\leftmark\hbox{}}\def\@evenfoot{}
\def\sectionmark##1{}\def\subsectionmark##1{}}

\oddsidemargin=\evensidemargin
\newcounter{sectionc}\newcounter{subsectionc}\newcounter{subsubsectionc}
\renewcommand{\section}[1] {\vspace{12pt}\addtocounter{sectionc}{1} 
\setcounter{subsectionc}{0}\setcounter{subsubsectionc}{0}\noindent 
        {\tenbf\thesectionc. #1}\par\vspace{5pt}}
\renewcommand{\subsection}[1] {\vspace{12pt}\addtocounter{subsectionc}{1} 
        \setcounter{subsubsectionc}{0}\noindent 
        {\bf\thesectionc.\thesubsectionc. {\kern1pt \bfit #1}}\par\vspace{5pt}}
\renewcommand{\subsubsection}[1] {\vspace{12pt}\addtocounter{subsubsectionc}{1}
        \noindent{\tenrm\thesectionc.\thesubsectionc.\thesubsubsectionc.
        {\kern1pt \tenit #1}}\par\vspace{5pt}}

\newcounter{appendixc}
\newcounter{subappendixc}[appendixc]
\newcounter{subsubappendixc}[subappendixc]
\renewcommand{\thesubappendixc}{\Alph{appendixc}.\arabic{subappendixc}}
\renewcommand{\thesubsubappendixc}
        {\Alph{appendixc}.\arabic{subappendixc}.\arabic{subsubappendixc}}

\renewcommand{\appendix}[1] {\vspace{12pt}
        \refstepcounter{appendixc}
        \setcounter{figure}{0}
        \setcounter{table}{0}
        \setcounter{lemma}{0}
        \setcounter{theorem}{0}
        \setcounter{corollary}{0}
        \setcounter{definition}{0}
        \setcounter{equation}{0}
        \renewcommand{\thefigure}{\Alph{appendixc}.\arabic{figure}}
        \renewcommand{\thetable}{\Alph{appendixc}.\arabic{table}}
        \renewcommand{\theappendixc}{\Alph{appendixc}}
        \renewcommand{\thelemma}{\Alph{appendixc}.\arabic{lemma}}
        \renewcommand{\thetheorem}{\Alph{appendixc}.\arabic{theorem}}
        \renewcommand{\thedefinition}{\Alph{appendixc}.\arabic{definition}}
        \renewcommand{\thecorollary}{\Alph{appendixc}.\arabic{corollary}}
        \renewcommand{\theequation}{\Alph{appendixc}.\arabic{equation}}
        \noindent{\tenbf Appendix \theappendixc #1}\par\vspace{5pt}}
\newcommand{\subappendix}[1] {\vspace{12pt}
        \refstepcounter{subappendixc}
        \noindent{\bf Appendix \thesubappendixc. {\kern1pt \bfit #1}}
        \par\vspace{5pt}}
\newcommand{\subsubappendix}[1] {\vspace{12pt}
        \refstepcounter{subsubappendixc}
        \noindent{\rm Appendix \thesubsubappendixc. {\kern1pt \tenit #1}}
        \par\vspace{5pt}}

\newcommand{\textlineskip}{\baselineskip=13pt}
\newcommand{\smalllineskip}{\baselineskip=10pt}

\def\eightcirc{
\begin{picture}(0,0)
\put(4.4,1.8){\circle{6.5}}
\end{picture}}
\def\eightcopyright{\eightcirc\kern2.7pt\hbox{\eightrm c}} 

\newcommand{\copyrightheading}[1]
        {\vspace*{-2.5cm}\smalllineskip{\flushleft
        {\footnotesize International Journal of Modern Physics D, #1}\\
        {\footnotesize $\eightcopyright$\, World Scientific Publishing
         Company}\\
         }}

\newcommand{\publisher}[2]{{\begin{center}\footnotesize\smalllineskip 
        Received #1\\
        Revised #2
        \end{center}
        }}

\def\abstracts#1#2#3{{
        \centering{\begin{minipage}{4.5in}\baselineskip=10pt\footnotesize
        \parindent=0pt #1\par 
        \parindent=15pt #2\par
        \parindent=15pt #3
        \end{minipage}}\par}} 

\renewenvironment{thebibliography}[1]
        {\frenchspacing
         \ninerm\baselineskip=11pt
         \begin{list}{\arabic{enumi}.}
        {\usecounter{enumi}\setlength{\parsep}{0pt}     
         \setlength{\leftmargin 12.7pt}{\rightmargin 0pt} 
         \setlength{\itemsep}{0pt} \settowidth
        {\labelwidth}{#1.}\sloppy}}{\end{list}}

\newcounter{itemlistc}
\newcounter{romanlistc}
\newcounter{alphlistc}
\newcounter{arabiclistc}

\newcommand{\fcaption}[1]{
        \refstepcounter{figure}
        \setbox\@tempboxa = \hbox{\footnotesize Fig.~\thefigure. #1}
        \ifdim \wd\@tempboxa > 5in
           {\begin{center}
        \parbox{5in}{\footnotesize\smalllineskip Fig.~\thefigure. #1}
            \end{center}}
        \else
             {\begin{center}
             {\footnotesize Fig.~\thefigure. #1}
              \end{center}}
        \fi}

\newcommand{\tcaption}[1]{
        \refstepcounter{table}
        \setbox\@tempboxa = \hbox{\footnotesize Table~\thetable. #1}
        \ifdim \wd\@tempboxa > 5in
           {\begin{center}
        \parbox{5in}{\footnotesize\smalllineskip Table~\thetable. #1}
            \end{center}}
        \else
             {\begin{center}
             {\footnotesize Table~\thetable. #1}
              \end{center}}
        \fi}

\def\@citex[#1]#2{\if@filesw\immediate\write\@auxout
        {\string\citation{#2}}\fi
\def\@citea{}\@cite{\@for\@citeb:=#2\do
        {\@citea\def\@citea{,}\@ifundefined
        {b@\@citeb}{{\bf ?}\@warning
        {Citation `\@citeb' on page \thepage \space undefined}}
        {\csname b@\@citeb\endcsname}}}{#1}}

\newif\if@cghi
\def\cite{\@cghitrue\@ifnextchar [{\@tempswatrue
        \@citex}{\@tempswafalse\@citex[]}}
\def\citelow{\@cghifalse\@ifnextchar [{\@tempswatrue
        \@citex}{\@tempswafalse\@citex[]}}
\def\@cite#1#2{{$\null^{#1}$\if@tempswa\typeout
        {IJCGA warning: optional citation argument 
        ignored: `#2'} \fi}}

\def\pmb#1{\setbox0=\hbox{#1}
        \kern-.025em\copy0\kern-\wd0
        \kern.05em\copy0\kern-\wd0
        \kern-.025em\raise.0433em\box0}

\def\fnt#1#2{\footnotetext{\kern-.3em
        {$^{\mbox{\scriptsize #1}}$}{#2}}}

\def\fpage#1{\begingroup
\voffset=.3in
\thispagestyle{empty}\begin{table}[b]\centerline{\footnotesize #1}
        \end{table}\endgroup}

\def\runninghead#1#2{\pagestyle{myheadings}
\markboth{{\protect\footnotesize\it{\quad #1}}\hfill}
{\hfill{\protect\footnotesize\it{#2\quad}}}}
\font\tenrm=cmr10
\font\tenit=cmti10 
\font\tenbf=cmbx10
\font\bfit=cmbxti10 at 10pt
\font\ninerm=cmr9

\font\eightrm=cmr8

\def\qed{\hbox{${\vcenter{\vbox{                        
   \hrule height 0.4pt\hbox{\vrule width 0.4pt height 6pt
   \kern5pt\vrule width 0.4pt}\hrule height 0.4pt}}}$}}

\begin{document}

\runninghead{Homothetic Wyman Spacetimes}{Homothetic Wyman Spacetimes}

\normalsize\textlineskip
\thispagestyle{empty}
\setcounter{page}{1}

\copyrightheading{Vol. 0, No. 0 (1997) 000--000}

\vspace*{0.88truein}

\begin{flushright}
\baselineskip=12pt
CERN--TH/97--166\\
\tt gr-qc/9707037
\end{flushright}

\fpage{1}
\centerline{\bf HOMOTHETIC WYMAN SPACETIMES}
\vspace*{0.37truein}
\centerline{\footnotesize M.A. Clayton}
\vspace*{0.015truein}
\centerline{\footnotesize\it CERN--Theory Division}
\baselineskip=10pt
\centerline{\footnotesize\it CH--1211 Geneva 23, Switzerland}
\vspace*{0.225truein}
\publisher{(received date)}{(revised date)}

\vspace*{0.21truein}
\abstracts{The time--dependent, spherically symmetric, Wyman sector of the Unified Field Theory is shown to be equivalent to a self--gravitating scalar field with a positive--definite, repulsive, self--interaction potential.
A homothetic symmetry is imposed on the fundamental tensor, and the resulting autonomous system is numerically integrated.
Near the critical point (between the collapsing and non-collapsing spacetimes) the system displays an approximately periodic alternation between collapsing and dispersive epochs.}{}{}

{\section{Introduction}\label{sect:Intro}}
\vspace*{-0.5pt}
\noindent
The Unified Field Theory (UFT)\cite{Einstein:1945,Einstein+Straus:1946} has held varying degrees of interest over the past $50$ years, however few would argue that the the essential idea (that of describing more physics by the ``geometric'' part of the field equations) was unwarranted.
One of the interesting features of such theories is the inability to identify a purely gravitational sector; gravitation is enfolded in a larger structure that was originally intended to describe both gravitational and electro-magnetic phenomena.
This is, of course, the reason for the name ``Unified Field theory'', and inevitably leads to a theories for which it is difficult even to formulate a strong equivalence principle.

Recently there has been some interest in claims that a reinterpretation of the UFT, the nonsymmetric gravitational theory (NGT),\cite{Moffat:1979,Moffat:1990} avoids singularity formation in collapse scenarios.\cite{Moffat:1995,Moffat+Sokolov:1995b}
This is motivated by the study\cite{Cornish+Moffat:1994,Cornish+Moffat:1994b,Cornish:1994b} of the static, spherically symmetric, Wyman solution.\cite{Wyman:1950,Bonnor:1951}
Claims that NGT should possess some repulsive force that will overcome the attractive force of Einstein gravity in strong field regimes and thereby prevent collapse to a black hole, have also appeared in the literature.\cite{Moffat:1995c} 
This drew counterarguments, taking the form of `thin shell' results\cite{Dobrowolski+Koc:1996} and the contention that for initial data that is `close enough' to a GR configuration the formation of a black hole is to be expected.\cite{Burko+Ori:1995}

In this work we add somewhat to the debate by mapping the Wyman sector field equations to those of General Relativity coupled to a self-interacting scalar field (referred to as the Einstein--Klein Gordon (E--KG) system) in Section~2, showing explicitly that the potential is indeed repulsive with respect to the centre of symmetry.
This shows up as explicit coordinate dependence in the potential, allowed in this case by the lack or separate conservation laws for the matter and gravitational fields (reflecting the lack of a strong equivalence principle).
We therefore expect to see some signs of a slowing of the collapse of a spherically--symmetric matter distribution, and indeed, we show that for the simplified spacetimes with a homothetic symmetry considered in Section~3 (mimicking the results on collapsing self--similar and homothetic E--KG spacetimes\cite{Brady:1995} which may be important for the near-critical scalar field collapse that has appeared in numerical investigations\cite{Choptuik:1993,Evans+Coleman:1994}), we see just such an effect.
In this way we hope to gain some insight into what to expect from the more general numerical collapse problem which will clarify this issue further.\cite{Clayton+Demopoulos+Legare:1997}

Perturbative analyses have shown that the original UFT/NGT field equations suffer from bad global asymptotics.\cite{Damour+Deser+McCarthy:1992,Damour+Deser+McCarthy:1993}
The most straightforward alteration of the action that avoids this criticism\cite{Moffat:1994,Moffat:1995b,Clayton:1995} is unstable near \textit{any} GR configuration,\cite{Clayton:1996,Clayton:1995b} and is therefore untenable as a physical theory.
Nevertheless, the spherically symmetric Wyman sector considered alone suffers from none of these flaws, and has remained essentially unchanged throughout these alterations.
Therefore we expect that when an action for NGT is discovered that is free of these problems (one such proposal exists\cite{Moffat:1996b}), the Wyman sector discussed herein will describe at least part of the spherically symmetric sector of the theory.

{\section{The Wyman Equations}\label{sect:reduction}}
\vspace*{-0.5pt}
\noindent
To begin, we will review the field equations for a spherically symmetric, self--gravitating, self--interacting scalar field in Bondi coordinates.
This will allow a straightforward map of the Wyman equations to this E--KG system with a specific choice of interaction potential.
Not only does this simplify the analysis of the field equations, but many results from E--KG may be ported directly to the Wyman system (the identification of a horizon, causal structure, mass function, \textit{etc.}).

\subsection{Spherically Symmetric, Self--Interacting, Scalar Field}
\label{sect:EKG}
\vspace*{-0.5pt}
\noindent
Generalising the development of Christodoulou\cite{Christodoulou:1986} (using the same partially null frame) to include the self--interacting scalar field, we consider the spherically symmetric spacetime metric in Bondi coordinates:
\begin{equation}\label{eq:metric}
\begin{split}
\mathrm{g}^{\textsc{e}}=&\mathrm{e}^{2\nu}du\otimes du
+\mathrm{e}^{\nu+\lambda}(du\otimes dr+dr\otimes du)
-r^2d\theta \otimes d\theta
-r^2\sin^2(\theta)d\phi \otimes d\phi,
\end{split}
\end{equation}
where $u$ is a null coordinate which is constant on the future light cone of all points on the central world line, and $r$ has been chosen so that spherical shells on these surfaces have area $4\pi r^2$.
Furthermore we introduce the partially null frame
\begin{gather}
e_n:=\mathrm{e}^{-\nu}\partial_u
-\tfrac{1}{2}\mathrm{e}^{-\lambda}\partial_r,\quad
e_l:=\mathrm{e}^{-\lambda}\partial_r,\quad
e_\theta:=r^{-1}\partial_\theta,\quad
e_\phi:=(r\sin(\theta))^{-1}\partial_\phi,
\end{gather}
in which the nonzero metric components are $\mathrm{g}^{\textsc{e}}_{nl}=\mathrm{g}^{\textsc{e}}_{ln}=-\mathrm{g}^{\textsc{e}}_{\theta\theta}=-\mathrm{g}^{\textsc{e}}_{\phi\phi}=1$.
We will also make use of the standard definitions\cite{Christodoulou:1986} $f:=\mathrm{e}^{\nu-\lambda}$, $g:=\mathrm{e}^{\nu+\lambda}$, the mass function $m:=\tfrac{r}{2}(1-\mathrm{e}^{-2\lambda})=\tfrac{r}{2}(1-f/g)$, and the derivative along the ingoing radial null lines $D:=\partial_u-\tfrac{1}{2}f\partial_r$.

The standard interacting scalar field Lagrangian ($V[\psi]\ge0$)
\begin{subequations}
\begin{equation}
\mathcal{L}:=\tfrac{1}{2}\mathrm{g}^{\mu\nu}\partial_\mu[\psi]\partial_\nu[\psi]-V[\psi]
=\mathrm{e}^{-(\nu+\lambda)}\partial_r[\psi]D[\psi]-V[\psi],
\end{equation}
results in the stress--energy tensor for the scalar field 
$T_{\mu\nu}=\partial_\mu[\psi]\partial_\nu[\psi]
-\mathrm{g}_{\mu\nu}\mathcal{L}$, which has frame components
\begin{equation}
T_{nn}=\mathrm{e}^{-2\nu}(D[\psi])^2,\quad
T_{ll}=\mathrm{e}^{-2\lambda}(\partial_r[\psi])^2,\quad
T_{nl}=T_{ln}=V[\psi],\quad
T_{\theta\theta}=T_{\phi\phi}=\mathcal{L}.
\end{equation}
\end{subequations}
The frame components of the Einstein tensor
\begin{subequations}\label{eq:GR equations}
\begin{align}
\label{eq:GR equations:ll}
G^{\textsc{e}}_{ll}=&R^{\textsc{e}}_{ll}=
\tfrac{2}{r}\mathrm{e}^{-2\lambda}\partial_r[\nu+\lambda],\\
\label{eq:GR equations:nl}
G^{\textsc{e}}_{nl}=&R^{\textsc{e}}_{\theta\theta}=
-\tfrac{1}{r^2}\mathrm{e}^{-(\nu+\lambda)}
\bigl(\partial_r[r\mathrm{e}^{\nu-\lambda}]-\mathrm{e}^{\nu+\lambda}\bigr),\\
\label{eq:GR equations:nn}
G^{\textsc{e}}_{nn}=&R^{\textsc{e}}_{nn}=
\tfrac{1}{2r}\mathrm{e}^{-2\lambda}\partial_r[\nu+\lambda]
-\tfrac{2}{r}\mathrm{e}^{-(\nu+\lambda)}\partial_u[\lambda],\\
\label{eq:GR equations:tt}
G^{\textsc{e}}_{\theta\theta}=&R^{\textsc{e}}_{nl}=
-\mathrm{e}^{-(\nu+\lambda)}\partial_r\bigl[\partial_u[\nu+\lambda]
-\mathrm{e}^{\nu-\lambda}\partial_r[\nu]\bigr]
+\tfrac{1}{r}\mathrm{e}^{-2\lambda}\partial_r[\nu-\lambda],
\end{align}
\end{subequations}
complete the description of Einstein's equations ($16\pi\mathrm{G}=\mathrm{c}=1$): $G^{\textsc{e}}_{\mu\nu}=\tfrac{1}{2}T_{\mu\nu}$. 

From the $ll$ and $nl$ field equations respectively, we find
\begin{align}\label{eq:GR diff}
\partial_r[\ln(g)]=&\tfrac{1}{4}r(\partial_r[\psi])^2,\quad
\partial_r[rf]=\bigl(1-\tfrac{1}{2}r^2V[\psi]\bigr)g,
\end{align}
the former leading to the integral solution (assuming that $\psi\sim\tfrac{1}{r}$ or faster as $r\rightarrow\infty$)
\begin{subequations}\label{eq:GR components}
\begin{equation}\label{eq:g solution}
g=\exp\Bigl[-\frac{1}{4}\int_r^\infty s(\partial_s[\psi(s)])^2\,ds\Bigr],
\end{equation}
indicating that $g$ is a nondecreasing function of $r$ on each surface of constant $u$, and since we have fixed $g\rightarrow 1$ as $r\rightarrow\infty$ for asymptotic flatness: $0\le g\le 1$.
The integral for $f$ is 
\begin{equation}\label{eq:f solution}
f=\overline{g}+f^\prime,\quad
\text{where}\quad
f^\prime:=\frac{1}{r}\int_r^\infty g(s)\frac{s^2}{2}V[\psi](s)\,ds. 
\end{equation}
\end{subequations}
Here we have introduced the radial average by $\overline{g}:=\tfrac{1}{r}\int_0^rg(s)\,ds$, and since $g$ is a nondecreasing function of $r$ we find that $\overline{g}\le g$ (using $g(u,r)\le g(u,r_{\text{max}})$ for $r\in[0,r_{\text{max}}]$ in the integral for $\overline{g}$).
We have also assumed that $f\rightarrow 1$ as $r\rightarrow\infty$, and that $\psi$ and the potential $V[\psi]$ are sufficiently well--behaved so that $f^\prime\rightarrow 0$ as $r\rightarrow\infty$.
Note that the presence of the potential results in $f^\prime\ge 0$, and therefore $f\ge \overline{g}$; allowing the possibility that $f>g$  and therefore $m<0$.

Since ingoing, radial, null geodesics are determined from 
\begin{equation}\label{eq:IRNG}
\partial_u[\chi(u;r)]=-\tfrac{1}{2}f\bigl(u,\chi(u;r)\bigr),
\end{equation}
with solution $\chi(u;r_0)=r_0-\tfrac{1}{2}\int_{u_0}^uf\bigl(s,\chi(s;r_0)\bigr)\,ds$, the local magnitude of $f$ controls the `rate' at which the geodesic is proceeding towards $r=0$.
Defining the displacement vector between two nearby geodesics as $\zeta(u;r_0):=\partial_{r_0}[\chi(u;r_0)]$, we derive the result that $\zeta(u;r_0)=\exp\bigl[-\tfrac{1}{2}\int_{u_0}^u\partial_r[f]\bigl(s,\chi(s;r_0)\bigr)\bigr]$, and we see that the local focusing of the ingoing radial null geodesics is governed by 
\begin{equation}
\partial_r[f]=\tfrac{1}{r}(g-f)
=\tfrac{1}{r}(g-\overline{g}-f^\prime).
\end{equation}
In the absence of the potential term we know that $f=\overline{g}\le g$ and so the geodesics are always focused towards $r=0$ (since then $\zeta\le 1$), however if $f>g$ then defocusing may occur.

It will also be useful to have an energy conservation equation; using $\partial_u[m]=r\mathrm{e}^{-2\lambda}\partial_u[\lambda]$ and~\eqref{eq:GR components}, we have $\partial_r[m]=\tfrac{r}{2}\mathrm{e}^{-2\lambda}\partial_r[\nu+\lambda]+\tfrac{1}{4}\mathrm{e}^{-(\nu+\lambda)}r^2V[\psi]$ and the $nn$ equation may be seen to be an evolution equation for the mass function along ingoing radial null geodesics
\begin{equation}\label{eq:mass equation}
\tfrac{2}{r^2}gD[m]=-\tfrac{1}{2}(D[\psi])^2-\tfrac{1}{4}fV[\psi].
\end{equation}

Inserting~\eqref{eq:GR diff} into the $G^{\text{\textsc{e}}}_{\theta\theta}$ component of the Einstein tensor yields
\begin{subequations}
\begin{equation}\label{eq:theta theta}
G^{\text{\textsc{e}}}_{\theta\theta}
=\tfrac{1}{2}g^{-1}\partial_r[\psi]D[\psi]-\tfrac{1}{2}V[\psi]
-\tfrac{1}{4}r\partial_r\bigl[V[\psi]\bigr]
-\tfrac{1}{4}r\partial_r[\psi]\nabla^2[\psi],
\end{equation}
which is equivalent to the remaining Einstein equation $G^{\text{\textsc{e}}}_{\theta\theta}=\tfrac{1}{2}T_{\theta\theta}$ only if
\begin{equation}\label{eq:cond1}
\partial_r[\psi]\nabla^2[\psi]
+\partial_r\bigl[V[\psi]\bigr]=0,
\end{equation}
\end{subequations}
where the covariant wave operator on $\psi$ is given by
\begin{equation}
\nabla^2[\psi]:=\mathrm{g}^{\mu\nu}\nabla_\mu\nabla_\nu[\psi]
=\tfrac{2}{r}g^{-1}D\bigl[\partial_r[r\psi]\bigr]-g^{-1}\partial_r[f]\partial_r[\psi].
\end{equation}
The equation of motion for $\psi$ derived from the action via the variational principle
\begin{equation}\label{eq:psi equation}
\nabla^2[\psi]+\delta_\psi\bigl[V[\psi]\bigr]=0,
\end{equation}
combined with~\eqref{eq:cond1} requires that $\partial_r\bigl[V[\psi]\bigr]=\delta_\psi\bigl[V[\psi]\bigr]\partial_r[\psi]$, which is true provided there is no explicit dependence on the radial coordinate in the potential.
We will have cause to re-examine this in the following section.

\subsection{The Wyman Equations}
\vspace*{-0.5pt}
\noindent
We will avoid giving the details of the computation of the field equations (the more general reduction of the system to $(1+1)$--dimensions will appear elsewhere\cite{Clayton:1997b}) instead quoting results derived using an existing method.\cite{Clayton:1995}
The field equations may be derived from the action\cite{Legare+Moffat:1995,Clayton:1995}
\begin{subequations}
\begin{equation}
S=\int d^4x\,\sqrt{-\mathrm{g}}\Bigl[
-\mathrm{g}^{\mu\nu}R^{\text{\textsc{ns}}}_{\mu\nu}
-\mathrm{g}^{\mu\nu}\partial_{[\mu}W_{\nu]}
+\tfrac{3}{8}\mathrm{g}^{\mu\nu}W_\mu W_\nu
+\mathrm{l}^\mu\Gamma_\mu
+\tfrac{1}{4}\mu^2\mathrm{g}^{[\mu\nu]}\mathrm{g}_{[\mu\nu]}
\Bigr],
\end{equation}
where 
\begin{equation}
R^{\text{\textsc{ns}}}_{\mu\nu}:=
\partial_\alpha\bigl[\Gamma^\alpha_{\mu\nu}\bigr]
-\tfrac{1}{2}\partial_\nu\bigl[\Gamma^\alpha_{\mu\alpha}\bigr]
-\tfrac{1}{2}\partial_\mu\bigl[\Gamma^\alpha_{\alpha\mu}\bigr]
+\Gamma^\alpha_{\mu\nu}\Gamma^\beta_{(\alpha\beta)}
-\Gamma^\alpha_{\beta\nu}\Gamma^\beta_{\mu\alpha}.
\end{equation}
\end{subequations}
The Wyman equations are those that result when the spherically symmetric ansatz (the functions $\gamma$, $\delta$, $\alpha$, $\beta$, and $f$ are functions of $t$ and $r$ only)
\begin{equation}\label{eq:Wyman}
\begin{split}
\mathrm{g}=&\gamma dt\otimes dt
+\delta\bigl(dt\otimes dr+dr\otimes dt\bigr)
-\alpha dr\otimes dr\\ 
&\quad-\beta \bigl(d\theta\otimes d\theta+\sin^2(\theta)d\phi\otimes d\phi\bigr)
+f\sin(\theta)\bigl(d\theta\otimes d\phi-d\phi\otimes d\theta\bigr),
\end{split}
\end{equation}
for the fundamental tensor is inserted into the general field equations, and have appeared various places in the literature.\cite{Wyman:1950,Bonnor:1951,Pant:1975,Tonnelat:1982}

Since the $t$--$r$ sector of~\eqref{eq:Wyman} is symmetric, we can once more make use of Bondi coordinates, and furthermore parameterise the antisymmetric part of the Wyman tensor using a dimensionless angle $\psi$ as
\begin{subequations}\label{eq:fundamental tensor}
\begin{equation}
\begin{split}
\mathrm{g}=&\mathrm{e}^{2\nu}du\otimes du
+\mathrm{e}^{\nu+\lambda}(du\otimes dr+dr\otimes du)\\
&-r^2\cos(\psi)\bigl
(d\theta \otimes d\theta+\sin^2(\theta)d\phi \otimes d\phi\bigr)
-r^2\sin(\psi)\sin(\theta)(d\theta\otimes d\phi-d\phi\otimes d\theta),
\end{split}
\end{equation}
which becomes
\begin{equation}\label{eq:ft frame}
\mathrm{g}=\theta^n\otimes\theta^l+\theta^l\otimes\theta^n
-\cos(\psi)\bigl(\theta^\theta\otimes\theta^\theta
+\theta^\phi\otimes\theta^\phi\bigr)
-\sin(\psi)\bigl(\theta^\theta\otimes\theta^\phi
-\theta^\phi\otimes\theta^\theta\bigr),
\end{equation}
\end{subequations}
in the partially null frame introduced in Section~2.1.
This parameterisation is particularly convenient since $\sqrt{-\mathrm{g}}=\sqrt{-\mathrm{g}^{\textsc{e}}}$ (\textit{i.e.}, is identical to the volume element of the metric~\eqref{eq:metric}) and the components of the inverse of~\eqref{eq:ft frame} are identical to that of~\eqref{eq:ft frame} with $\psi\rightarrow -\psi$ (the angular components of~\eqref{eq:ft frame} have the form of a rotation matrix).

The field equations (in the form $R^{\textsc{ns}}_{\mu\nu}=0$\cite{Clayton:1995}) may be found by straightforward computation, and are related to the results of the last section by
\begin{subequations}\label{eq:NGT field equations}
\begin{gather}
R^{\textsc{ns}}_{ll}=R^{\textsc{e}}_{ll}
-\tfrac{1}{2}\mathrm{e}^{-2\lambda}(\partial_r[\psi])^2=0,\\
\cos(\psi)R^{\textsc{ns}}_{\theta\theta}-\sin(\psi)R^{\textsc{ns}}_{\theta\phi}
=R^{\textsc{e}}_{\theta\theta}
-\tfrac{1}{2}V[\psi]=0,\\
R^{\textsc{ns}}_{nn}=R^{\textsc{e}}_{nn}
-\tfrac{1}{2}\mathrm{e}^{-2\nu}(D[\psi])^2=0,\\
\label{eq:skew}
\sin(\psi)R^{\textsc{ns}}_{\theta\theta}+\cos(\psi)R^{\textsc{ns}}_{\theta\phi}
=\tfrac{1}{2}\bigl(\nabla^2[\psi]+\delta_\psi\bigl[V[\psi]\bigr]\bigr)=0,\\
\label{eq:odd}
R^{\textsc{ns}}_{nl}=R^{\textsc{e}}_{nl}
-\tfrac{1}{2}\mathrm{e}^{-(\nu+\lambda)}\partial_r[\psi]D[\psi]
+\tfrac{1}{4}\mu^2\sin^2(\psi)=0,
\end{gather}
\end{subequations}
where $f$ and $g$ are defined as in Section~2.1, and the potential is given by
\begin{equation}\label{eq:NGT potential}
V[\psi]=\tfrac{2}{r^2}\bigl(1-\cos(\psi)\bigr)
+\tfrac{1}{2}\mu^2\sin^2(\psi).
\end{equation}
For explicitness we give here the field equations in the form most useful for later considerations:
\begin{subequations}\label{eq:feqs}
\begin{gather}
\partial_r[\ln g]=\tfrac{1}{4}r\bigl(\partial_r[\psi]\bigr)^2,\\
\partial_r[rf]=\cos(\psi)g-\tfrac{1}{4}(\mu r)^2g\sin^2(\psi),\\
2rD\bigl[\partial_r[r\psi]\bigr]
-r^2\partial_r[f]\partial_r[\psi]
+2g\sin(\psi)+(\mu r)^2g\sin(\psi)\cos(\psi)=0.
\end{gather}
\end{subequations}
It is a rather laborious task to show that the Wyman solution\cite{Wyman:1950,Bonnor:1951,Cornish:1994} identically satisfies these field equations with $\mu=0$.

The first three equations of~\eqref{eq:NGT field equations} are clearly equivalent to the $ll$, $nl$ and $nn$ field equations of the E--KG system~\eqref{eq:GR equations} with the potential given by~\eqref{eq:NGT potential}, leading to~\eqref{eq:g solution},~\eqref{eq:f solution}, and~\eqref{eq:mass equation}.
In~\eqref{eq:skew} we find the field equation for $\psi$ as would be derived from the variational principle with the potential~\eqref{eq:NGT potential}.
The final field equation~\eqref{eq:odd} is somewhat odd looking since it would appear to be incompatible with both of~\eqref{eq:NGT potential} and the field equation $G^{\textsc{e}}_{\theta\theta}=\tfrac{1}{2}T_{\theta\theta}$.
However if we allow for the possibility that the potential depends explicitly on the radial coordinate, then the partial derivative of the potential splits up into two parts (the second term represents the derivative of the potential with respect to the explicit coordinate dependence only)
\begin{equation}
\partial_r\bigl[V[\psi]\bigr]=\delta_\psi\bigl[V[\psi]\bigr]\partial_r[\psi]
+\partial_r[V][\psi], 
\end{equation}
and instead of~\eqref{eq:theta theta} we have
\begin{equation}\label{eq:r dependent}
G^{\textsc{e}}_{\theta\theta}
=\tfrac{1}{2}g^{-1}\partial_r[\psi]D[\psi]
-\tfrac{1}{2}V[\psi]
-\tfrac{1}{4}r\partial_r[V][\psi]
-\tfrac{1}{4}r\partial_r[\psi]
\bigl(\nabla^2[\psi]+\delta_\psi\bigl[V[\psi]\bigr]\bigr).
\end{equation}
However by direct computation, we finds from~\eqref{eq:NGT potential} that
\begin{equation}
V[\psi]+\tfrac{1}{2}r\partial_r[V][\psi]=\tfrac{1}{2}\mu^2\sin^2(\psi),
\end{equation}
and therefore the field equations are consistent.

As advertised, the potential~\eqref{eq:NGT potential} acts (for non-zero $\psi$) as a repulsive central force; indeed we expect that it will overcome the attractive gravitational potential at small enough radii.
In the E--KG system, explicit coordinate dependence in the scalar potential is in conflict with the Bianchi identities; in flat spacetime leading to conservation laws of the form\cite{Goldstein:1980} $\partial\cdot T=-\partial\mathcal{L}$.
We stress though that what is playing the role of the Bianchi identity in UFT/NGT\cite{Lichnerowicz:1954,Legare+Moffat:1995} is \textit{not} the same as in GR.
For the E--KG system we had (for spherical symmetry) a field equation for the scalar field derived from the variational principle that was equivalent to the conservation laws (and therefore Einstein's field equations) provided there was no explicit dependence on the coordinates.
Here we have a (diffeomorphism invariant) system of equations that includes the scalar field equation that altogether satisfies the UFT/NGT Bianchi identities, but does \textit{not} satisfy the separate E--KG conservation laws.

For this work the operationally important point is that the functions that parameterise the symmetric part of the fundamental tensor are determined in exactly the same way from the scalar field as if it were a spherically--symmetric, E--KG field.
Furthermore, the field equation for $\psi$ is exactly that which would be determined from the variation of the scalar field action.
Therefore we can treat the system as E--KG, ignoring the fact that the potential has explicit coordinate dependence.

{\section{Self--Similar Solutions}\label{sect:homothetic}}
\vspace*{-0.5pt}
\noindent
Evidence for self--similarity in the numerical collapse of a scalar field was first noted by Choptuik,\cite{Choptuik:1993} and sparked much interest in the literature.
Since that time it seems accepted that it is a discrete self--similarity rather than the the continuous (or homothetic) self--similarity considered herein which is observed.\cite{Gundlach:1997}
Nevertheless we feel that there is something to be learned in examining the homothetic solutions: not only are the resulting equations much simpler to integrate, but the behaviour which results is reflecting the repulsive nature of the potential~\eqref{eq:NGT potential} and should be observable in more generic collapse scenarios.

We now make the ansatz\cite{Cahill+Taub:1971} that the fundamental tensor has a homothetic symmetry, that is, there exists a vector field $\zeta$ for which the fundamental tensor satisfies 
\begin{equation}\label{eq:homo}
\pounds_\zeta[\mathrm{g}]=2\mathrm{g}.
\end{equation}
It is straightforward to show that for the fundamental tensor~\eqref{eq:fundamental tensor} we have
\begin{equation}
\pounds_\zeta[\lambda]=0,\quad
\pounds_\zeta[\nu]=0,\quad
\pounds_\zeta[\psi]=0,
\end{equation}
where the similarity vector is
\begin{equation}
\zeta=u\partial_u+r\partial_r,
\end{equation}
and so all three functions must depend solely on the variable $x:=r/u$.
From the results of the last section we know that this is equivalent to considering a homothetic E--KG system with a nontrivial potential, restricted to the scale invariant case ($\kappa=0$) of Brady.\cite{Brady:1995}

Indeed, normally if one considers a scalar field potential and requires spacetime to have a homothetic symmetry, one ends up with the requirement that $\pounds_\zeta\bigl[V[\psi]\bigr]+2V[\psi]=0$, which for potentials that are polynomial in the scalar field, requires that $\psi=0$, and one is left with a trivial spacetime.
To see this we follow the argument in the appendix~A of Brady\cite{Brady:1995} and note that $\pounds_\zeta[R]_{\mu\nu}=0$ now implies that $\pounds_\zeta\bigl[\nabla_\mu[\psi]\nabla_\nu[\psi]-\mathrm{g}_{\mu\nu}V[\psi]\bigr]=0$.
Considering an angular component of this and using spherical symmetry and~\eqref{eq:homo} yields the result.
Using the potential~\eqref{eq:NGT potential} and $\pounds_\zeta[\psi]=0$, we satisfy this requirement due to the explicit dependence on the radial coordinate in the potential.
Therefore the existence of self--similar Wyman spacetimes is \textit{not} a trivial extension of the E--KG results, it is actually a rather surprising possibility that is wrapped up in the structure of the field equations.
(Note that we are considering $\mu=0$ throughout; the existence of this length scale would clearly contradict scale invariance.
We expect however that if the homothetic symmetry is relevant to the general collapse problem, then the generic features presented herein should also appear for $\mu\neq 0$ on distance scales $<\mu^{-1}$.\cite{Gundlach:1997})

Inserting the ansatz into the field equations~\eqref{eq:feqs} we find
\begin{subequations}\label{eq:x equations}
\begin{gather}
\partial_x[\ln g]=\tfrac{1}{4}x\bigl(\partial_x[\psi]\bigr)^2,\\
\partial_x[xf]=\cos(\psi)g,\\
\label{eq:x psi equation}
2x(x+\tfrac{1}{2}f)\partial_x^2[x\psi]
+x^2\partial_x[f]\partial_x[\psi]
-2g\sin(\psi)=0.
\end{gather}
\end{subequations}
It will be important later to determine the large $x$ asymptotic behaviour of asymptotically flat solutions of~\eqref{eq:x equations}.
Assuming that $f\rightarrow f_0$ and $g\rightarrow g_0$ (\textit{i.e.}, spacetime is flat in the asymptotic limit) we find that the asymptotically dominant contribution to~\eqref{eq:x psi equation} is 
\begin{equation}
x^3\partial_x^2[\psi]
+2x^2\partial_x[\psi]
-g_0\sin(\psi)\sim 0.
\end{equation}
If $\psi\rightarrow \psi_0\neq 0$, we can approximate $\sin(\psi)\sim\sin(\psi_0)+\cos(\psi_0)\tilde{\psi}$ where $\psi\rightarrow\psi_0+\tilde{\psi}$, and then taking $\lvert x\rvert=:g_0\cos(\psi_0)y$ and $\lambda:=\tilde{\psi}+\tan(\psi_0)$ we have
\begin{equation}
y^3\partial_y^2[\lambda]+2y^2\partial_y[\lambda]\mp\lambda\sim 0,
\end{equation}
where the upper sign is for $x>0$ and the lower for $x<0$.
Introducing $y=a/z^2$ it is easy to see that the solutions to this are Bessel functions, and once we requires that $\psi\rightarrow\psi_0$ we have the asymptotic form of the solutions
\begin{subequations}
\begin{align}
\psi_{x>0}\sim&\psi_0+\tan(\psi_0)\bigl(zK_1(z)-1\bigr)+CzI_1(z),\\
\label{eq:greater asymptote}
\psi_{x<0}\sim&\psi_0-\tan(\psi_0)\bigl(\tfrac{1}{2}\pi zY_1(z)+1\bigr)+CzJ_1(z),
\end{align}
\end{subequations}
where $C$ is an integration parameter.
As it turns out, the case where $\psi\rightarrow 0$ is equivalent to taking $\psi_0=0$ in these solutions.

Returning now to the system~\eqref{eq:x equations}, we introduce the alternate variables $Y:=f/g$, $Z:=x/f$, the coordinate ${t}=\ln\lvert x\rvert$ and the additional variable $P:=\dot{\psi}$ (an overdot indicating a derivative with respect to $t$), and find the autonomous system
\begin{subequations}\label{eq:first order}
\begin{align}
\dot{Z}=&Z\bigl(2-Y^{-1}\cos(\psi)\bigr),\\
\label{eq:NGT Y dot}
\dot{Y}=&\cos(\psi)-Y
-\tfrac{1}{4}YP^2
=\frac{2(Z+1)}{2Z+1}\bigl(\cos(\psi)-Y\bigr),\\
\dot{\psi}=&P,\\
\dot{P}=&
-Y^{-1}(2Z+1)^{-1}
\bigl[\bigl(2YZ+\cos(\psi)\bigr)P
-2\sin(\psi)\bigr].
\end{align}
\end{subequations}
where in~\eqref{eq:NGT Y dot} we have made use of the energy conservation equation~\eqref{eq:mass equation} which, assuming homothetic symmetry, may be written in the useful forms
\begin{equation}\label{eq:energy conservation}
P^2=\frac{4\bigl(1-Y^{-1}\cos(\psi)\bigr)}{2Z+1}
=-\frac{2}{Z+1}\partial_{t}[\ln Y].
\end{equation}

Note that $\{-\infty<{t}<\infty\}$ covers either $u<0$ or $0<u$, however the solutions are mapped into each other by ${t}\leftrightarrow -{t}$.
We will be considering the former case exclusively where spacetime is empty and flat for ${t}<1$ (so that $\psi=0$, $Y=1$ and $Z=x/f_i$ for some constant $f_i$).
By rescaling of the $u$ coordinate $f_0$ may be chosen at will; however this merely shifts ${t}\rightarrow{t}_0+{t}$, mapping each trajectory onto itself.
At ${t}=1$ the scalar field will be `turned on' (\textit{i.e.}, matched to a self--similar solution through the initial data) and the system evolved as far as possible towards ${t}=\infty$ ($u=0$).

\subsection{The Einstein--Klein Gordon Limit.}
\vspace*{-0.5pt}
\noindent
The equivalent set of equations for a non-interacting, self-gravitating, spherically symmetric, scalar field are equivalent to the $V[\psi]=0$ limit of~\eqref{eq:first order}:
\begin{subequations}\label{eq:GR fieldeqns}
\begin{align}
\label{eq:GR Zdot}
\dot{Z}=&Z(2-Y^{-1}),\\
\label{eq:GR Ydot}
\dot{Y}=&\frac{2(Z+1)}{2Z+1}(1-Y),\\
\dot{\psi}=&P,\quad
\dot{P}=
-Y^{-1}(2Z+1)^{-1}
\bigl[(2YZ+1)P\bigr],
\end{align}
\end{subequations}
and 
\begin{equation}\label{eq:GR psq}
P^2=\frac{4(1-Y^{-1})}{2Z+1}
=-\frac{2}{Z+1}\partial_{t}[\ln Y],
\end{equation}
replaces~\eqref{eq:energy conservation}.

Since a free scalar field has a positive mass function by construction (see~\eqref{eq:f solution} with $V[\psi]=0$) we know that $Y\le 1$, we furthermore confine ourselves to $Y\ge 0$ since $Y=0$ signals the presence of an apparent horizon.
With this information in hand, the requirement that~\eqref{eq:GR psq} be positive (and therefore $P$ real) results in $Z\le-\tfrac{1}{2}$.
Ingoing radial null geodesics may be reparameterised in terms of $x$ as $\chi(u;r)=u\chi(u;x)$ which, inserted into~\eqref{eq:IRNG} and integrated from $u=-1$ (corresponding to $x_1$) to $u$ gives
\begin{equation}\label{eq:u}
\ln\lvert u\rvert=-2\int_{x_1}^x dx^\prime\,
\bigl(f(x)+2x\bigr)^{-1},
\end{equation}
and with $f=x/Z$ and $Z<-\tfrac{1}{2}$ we see that all of the region of interest is accessible, and either $u\rightarrow 0$ while $Z\rightarrow\infty$ or $u\rightarrow u_0$ at the horizon $Z=-\tfrac{1}{2}$.

Taking the ratio of~\eqref{eq:GR Ydot} and~\eqref{eq:GR Zdot} we can write $Y=Y[Z]$ to find 
\begin{equation}\label{eq:ratio}
\frac{\partial Y}{\partial Z}=\frac{2(Z+1)}{Z(2Z+1)}\frac{Y(1-Y)}{2Y-1},
\end{equation}
which implies that trajectories may only pass through the lines $Z=-\tfrac{1}{2}$, $Y=0$ or $Y=-1$ at either $\bigl(-\tfrac{1}{2},0\bigr)$ or $\bigl(-\tfrac{1}{2},1\bigr)$, and must cross $Y=\tfrac{1}{2}$ perpendicularly unless $Z=-1$.
The point $\bigl(-1,\tfrac{1}{2}\bigr)$ is an unstable fixed point of~\eqref{eq:GR Zdot} and~\eqref{eq:GR Ydot}.\cite{Brady:1995}
(We refer to points in the $Z$--$Y$, and later $Z$--$X$, plane by their coordinates: $\bigl(-\tfrac{1}{2},0\bigr)$ refers to the point $Z=-\tfrac{1}{2}$, $Y=0$.)
Additionally, from the second form of~\eqref{eq:GR psq} $Y>0$ implies that $\dot{Y}<0$ for $-1<Z<-\tfrac{1}{2}$ and $\dot{Y}>0$ for $Z<-1$, and~\eqref{eq:GR Zdot} implies that $\dot{Z}<0$ for $\tfrac{1}{2}<Y<1$ and $\dot{Z}>0$ for $0<Y<\tfrac{1}{2}$.
Therefore trajectories that are matchable to a flat spacetime (only $C^1$) emanate from $\bigl(-\tfrac{1}{2},1\bigr)$ ending at either $\bigl(-\tfrac{1}{2},0\bigr)$ (a horizon), $\bigl(-1,\tfrac{1}{2}\bigr)$ (asymptotically approaching the fixed point), or travelling to infinite $Z$ with asymptotic form
\begin{equation}
Y\approx 1+Y_0\mathrm{e}^{-{t}},\quad
Z\approx Z_0\mathrm{e}^{t},
\end{equation}
which approaches a flat spacetime configuration.
Nothing need be said of $\psi$ since it does not directly enter into the equations for $Y$ and $Z$; however since $P$ is of a fixed sign throughout these trajectories, if $\psi=0$ initially it \textit{cannot} vanish again.
For a free scalar field this is irrelevant since a constant $\psi$ solution has vanishing energy momentum tensor.
This will clearly not be the case for the homothetic Wyman spacetimes.

It is useful that there exists an exact solution to this system,\cite{Roberts:1989} which we will reproduce here.
Integrating~\eqref{eq:ratio} results in 
\begin{subequations}\label{eq:GR exact soln}
\begin{equation}\label{eq:soln a}
Y(1-Y)=-\frac{A}{4}\frac{2Z+1}{Z^2},\quad\rightarrow\quad
Y=\frac{1}{2Z}\bigl(Z\pm\sqrt{Z^2+A(2Z+1)}\bigr),
\end{equation}
where $A>0$ is a parameter of integration chosen so that the critical solution corresponds to $A=A_c:=1$, sub-critical $0\le A<1$ and super-critical $A>1$.
Note that the $\pm$ sign in~\eqref{eq:soln a} refers to $\tfrac{1}{2}<Y<1$ and $0<Y<\tfrac{1}{2}$ respectively.
Inserting this into~\eqref{eq:GR Zdot} and requiring that $Z=-\tfrac{1}{2}$ at ${t}=0$ gives
\begin{equation}
Z=-\frac{1}{2}\frac{\mathrm{e}^{2{t}}}{A+\mathrm{e}^{2{t}}}
\Bigl(A+\sqrt{A(1-\mathrm{e}^{2{t}})+\mathrm{e}^{2{t}}}\Bigr).
\end{equation}
\end{subequations}
Note that super-critical solutions reach $\bigl(-\tfrac{1}{2},0\bigr)$ in a finite parameter time ${t}=\tfrac{1}{2}\ln\bigl(A/(A-1)\bigr)$.

\begin{figure}
\begin{center}
\includegraphics[scale=0.53 ,angle=90]{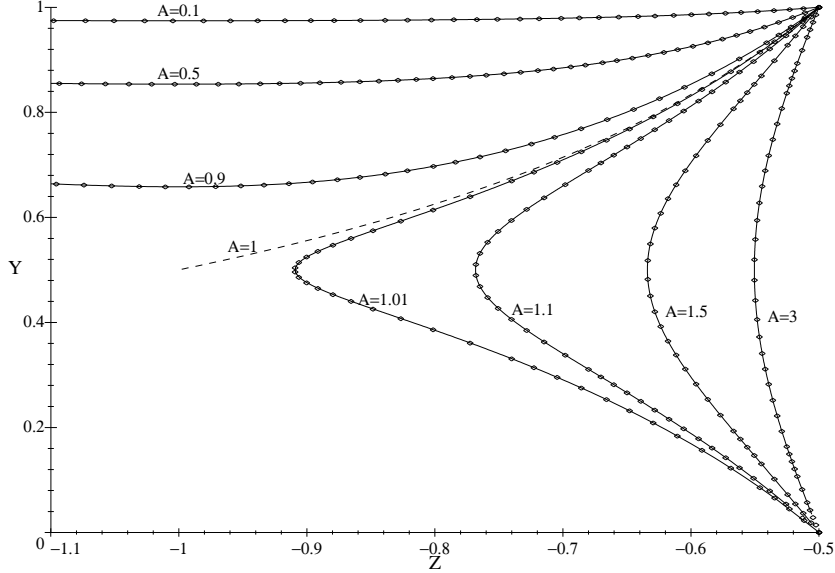}
\end{center}
\caption{
E--KG trajectories.  
The lines are calculated by numerically integrating the equations~\eqref{eq:GR fieldeqns}, whereas the diamonds are calculated via the exact solution~\eqref{eq:GR exact soln}.  
The dotted line indicates the critical solution which asymptotically approaches the fixed point at $\bigl(-1,\tfrac{1}{2}\bigr)$.
}
\label{fig:GR trajectories}
\end{figure}

In Figure~\ref{fig:GR trajectories} we compare this exact solution~\eqref{eq:GR exact soln} (indicated by diamonds) to that derived from numerically integrating the system of equations~\eqref{eq:GR fieldeqns} (solid lines), also indicating the exactly critical solution ($A=1$). 
In order to avoid the poor behaviour of $(2Z+1)^{-1}$ in~\eqref{eq:GR fieldeqns} we begin the numerically generated trajectories near $\bigl(-\tfrac{1}{2},1\bigr)$ by considering initial data from the exact solution~\eqref{eq:GR exact soln} Taylor expanded to a small parameter time $\Delta{t}$
\begin{equation}\label{eq:Taylor}
Y\approx 1-A\Delta{t},\quad
Z\approx-\tfrac{1}{2}(1+\Delta{t}),\quad
\psi\approx \pm 2\sqrt{A}\Delta{t},\quad
P\approx \pm 2\sqrt{A}.
\end{equation}
Note that the choice of sign is immaterial, merely replacing $\psi\leftrightarrow -\psi$ and $P\leftrightarrow -P$ along the entire trajectory.
The initial time $\Delta{t}$ was chosen by $A\Delta{t}=10^{-4}$ which, for the range of $A$ considered throughout this work, was sufficient for considering qualitative features of these spacetimes.
In generating all trajectories, we made use of the numerical ode package of Maple V Release 4;\cite{MapleVr4,Redfern:1996} although we have performed no quantitative error estimates, the agreement of the numerical versus exact results in Figure~\ref{fig:GR trajectories}, qualitative agreement with the analytic conditions derived, and the robustness of the results using the different integration schemes supplied by Maple provide confidence in the results.
The results presented were generated using the Livermore Stiff ODE solver (lsode) routine with default options.

\subsection{The Wyman Equations}
\vspace*{-0.5pt}
\noindent
As before we begin by noting that we must have $Y\ge 0$ in order to be considering a (part of) spacetime that is not in the interior of a black hole.
In this case however, we cannot claim that $Y\le 1$ since by~\eqref{eq:f solution} we need not have a positive--definite mass function.
However using the explicit form of the potential~\eqref{eq:NGT potential} (with $\mu=0$) we do find that $f=\tfrac{1}{r}\int_0^rdr\,\cos(\psi)g$, which allows the simple bound $\lvert f\rvert\le g$, and we need only consider spacetimes with $0\le Y\le 2$.
In spite of this, we shall see that all of the trajectories we consider will remain within the region $0\le Y\le 1$.

It is useful to consider the new variable $X:=Y/\cos(\psi)$, in terms of which~\eqref{eq:first order} takes on a form which is similar to~\eqref{eq:GR fieldeqns}
\begin{subequations}\label{eq:Z eqns}
\begin{align}
\label{eq:Z dot}
\dot{Z}=&Z\bigl(2-X^{-1}\bigr),\\
\label{eq:X dot}
\dot{X}=&
\frac{2(Z+1)}{2Z+1}\bigl(1-X\bigr)+XP\tan(\psi),\\
\label{eq:psi dot}
\dot{\psi}=&P,\\
\label{eq:P dot}
\dot{P}=&
-X^{-1}(2Z+1)^{-1}
\bigl[\bigl(2XZ+1\bigr)P
-2\tan(\psi)\bigr],
\end{align}
\end{subequations}
and 
\begin{equation}\label{eq:X psq}
P^2=\frac{4(1-X^{-1})}{2Z+1}
=-\frac{2}{Z+1}\partial_{t}[\ln X].
\end{equation}
The energy conservation equation is now identical to~\eqref{eq:GR psq}, and since there is no physical bound on $X$, we find that trajectories must be in one of the following regions:
\begin{equation}
\begin{cases}
\mathrm{i}  \quad & X<0,\quad Z>-\tfrac{1}{2}\\
\mathrm{ii} \quad & 0<X<1,\quad Z<-\tfrac{1}{2}\\
\mathrm{iii}\quad & X>1,\quad Z>-\tfrac{1}{2}\\
\end{cases}.
\end{equation}
From our initial conditions, we require that $\psi=0$ and $Y=1$, implying that trajectories of interest must begin at $\bigl(-\tfrac{1}{2},1\bigr)$.
For a suitably small interval in ${t}$ (where the $\tan(\psi)$ term in~\eqref{eq:X dot} has little effect) we will find approximately E--KG evolution, and therefore trajectories will flow into region~$\mathrm{ii}$.
Moreover, since we will never find that $\psi\rightarrow \bigl(n+\tfrac{1}{2}\bigr)\pi$ on any trajectory, we may also unambiguously identify $X=0$ as a horizon, and may therefore consider region~$\mathrm{ii}$ exclusively where again we see from~\eqref{eq:u} that the entire region of interest is accessible to incoming radial null geodesics.
The fixed point $\bigl(-1,\tfrac{1}{2}\bigr)$ that existed for the E--KG system may no longer be considered a fixed point since we have $P=\dot{\psi}=4\neq 0$ (note that $P=4$ also for the E--KG system, however there is no potential to feed back into the $(Z,Y)$ system; this means that $\psi\sim \psi_0+4t$ along this `fixed point').

Taking the ratio of~\eqref{eq:X dot} to~\eqref{eq:Z dot} we find
\begin{equation}
\frac{\partial X}{\partial Z}=\frac{2(Z+1)}{Z(2Z+1)}\frac{X(1-X)}{2X-1}
+\frac{X^2}{2X-1}\frac{P\tan(\psi)}{Z}.
\end{equation}
As for the E--KG system, trajectories may not cross $Z=-\tfrac{1}{2}$ or $X=0$ except at $\bigl(-\tfrac{1}{2},0\bigr)$ or $\bigl(-\tfrac{1}{2},1\bigr)$, however they may intersect $X=1$.
Trajectories that do so at $Z\neq -\tfrac{1}{2}$ must intersect with $P=0$ from~\eqref{eq:X psq}, so if we consider initial points at $X=1$, $Z\neq-\tfrac{1}{2}$ then we initially have $\psi=P=0$.
It is easy to see that the solution is given by $\psi=0$, $P=0$, $X=1$, and $Z=Z_0\mathrm{e}^{t}$ which is the flat spacetime solution.
From~\eqref{eq:ratio} we deduce that trajectories will pass through $X=\tfrac{1}{2}$ perpendicularly unless $P(Z+1)=2\tan(\psi)$, which implies that $\dot{P}=0$ and therefore that $\dot{X}=0$ and so are always perpendicular.

\begin{figure}
\centering
\subfigure[Wyman Trajectories in the $Z$--$X$ plane.]{\label{fig:all X}
\includegraphics[scale=0.28 ,angle=90]{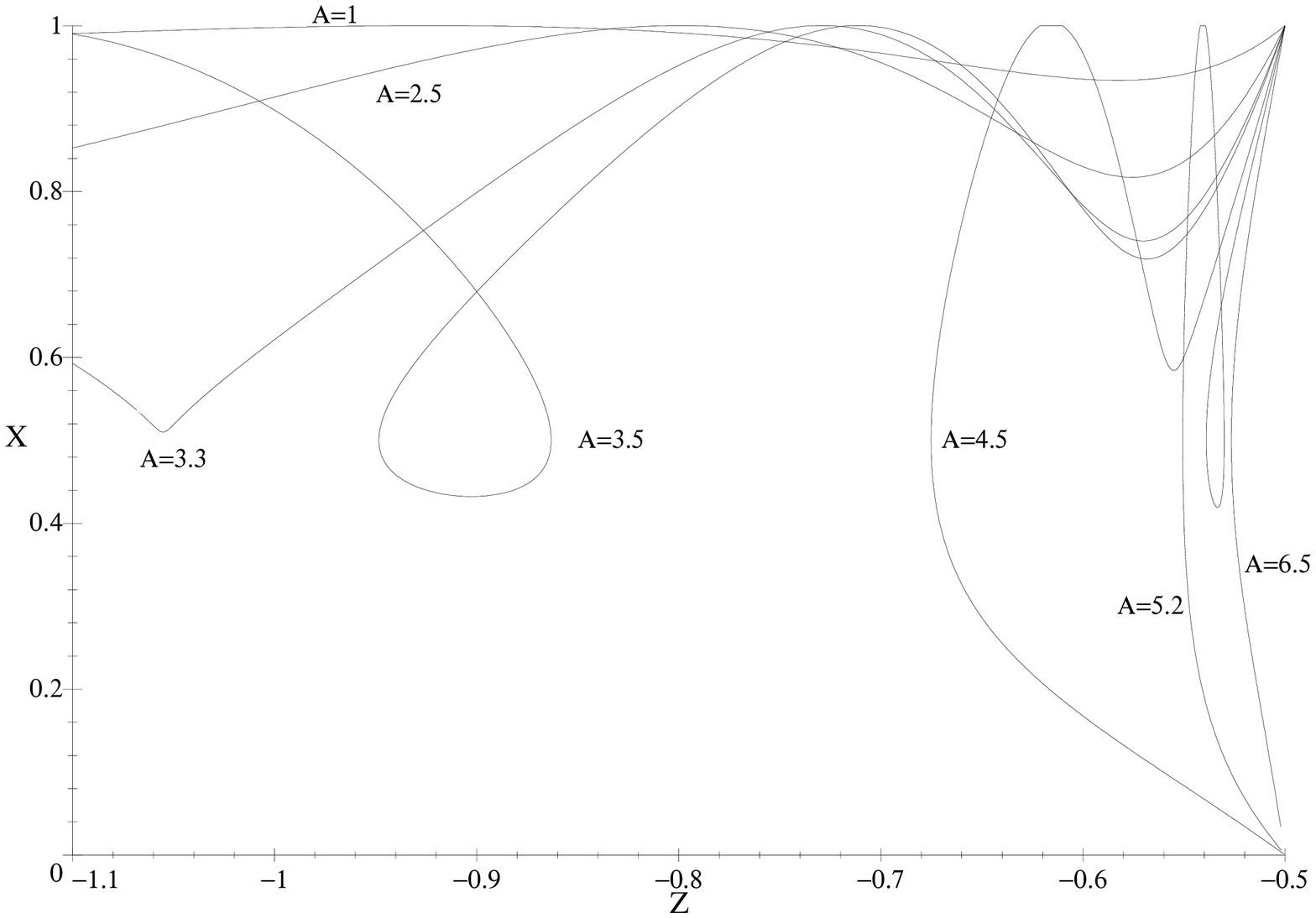}}
\hspace{0.3cm}
\subfigure[Wyman Trajectories in the $Z$--$Y$ plane.]{\label{fig:all Y}
\includegraphics[scale=0.28 ,angle=90]{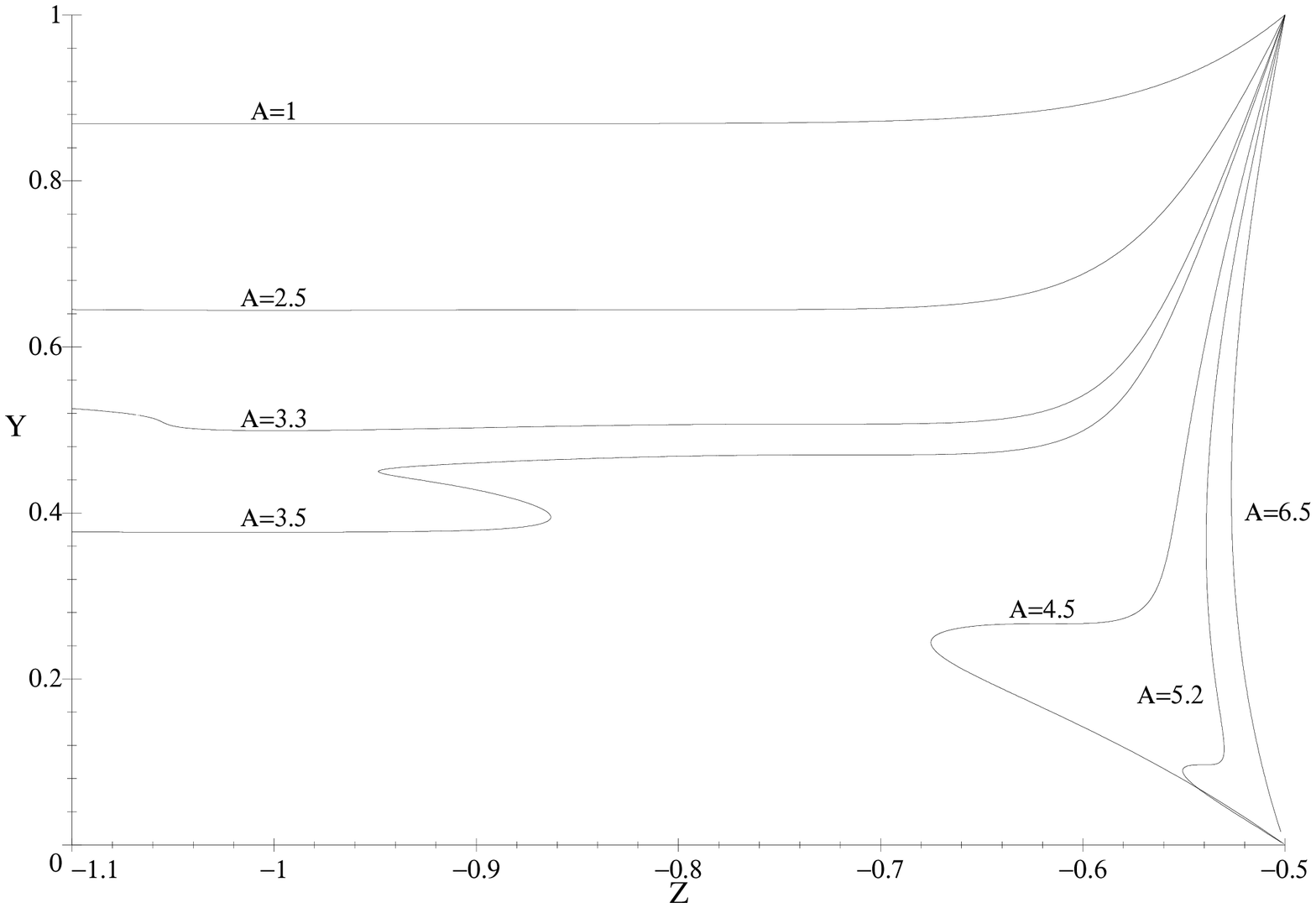}}
\label{fig:all}
\caption{A selection of the Wyman trajectories showing sub-critical ($A=1$ and $A=2.5$), sub-critical approaching ($A=3.3$) and crossing ($A=3.5$) the line $X=0.5$, super-critical on the second oscillation with the first oscillation not passing through $X=0.5$ ($A=4.5$) and passing through $X=0.5$ ($A=5.2$), and finally a `direct' super-critical trajectory ($A=6.5$).}
\label{fig:generic}
\end{figure}

\begin{figure}
\begin{center}
\includegraphics[scale=0.53 ,angle=90]{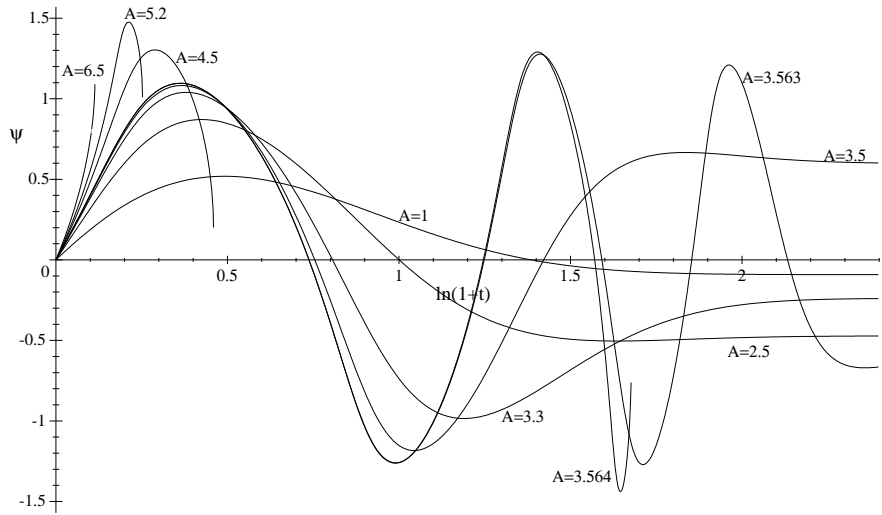}
\end{center}
\caption{The scalar field $\psi$ for the trajectories displayed in Figure~\ref{fig:generic} as well as the two near-critical trajectories in Figure~\ref{fig:near critical}; note the logarithmic time scale.
}
\label{fig:psi}
\end{figure}

We display a variety of trajectories emanating from $\bigl(-\tfrac{1}{2},1\bigr)$ in Figure~\ref{fig:generic}.
All sub-critical trajectories eventually approach $X=1$ which, as we see from Figure~\ref{fig:psi}, does \textit{not} in general correspond to $\psi=0$; instead their form corresponds to that given by~\eqref{eq:greater asymptote}.
However since the asymptotic form does \textit{not} correspond to a solution of~\eqref{eq:feqs}, we cannot match onto a flat spacetime solution at $u=0$ (as one could for the E--KG case, taking $\psi=\text{constant}$).
It seems likely that the trajectory would continue on through $u=0$, matching to the equivalent $u>0$ trajectory, effectively following the same paths as shown in Figure~\ref{fig:generic} back to $Y=1$, at which point we could match back onto Minkowski space.
We also find that there are no trajectories with $\psi\rightarrow\pm\pi/2$ (those which approach $\pm\tfrac{1}{2}\pi$ are repelled by the $\tan(\psi)$ term in~\eqref{eq:P dot} since $P\approx 0$) passing through $\bigl(-\tfrac{1}{2},0\bigr)$, so we conclude that there is always a horizon and need not continue the trajectory into region~$\mathrm{i}$.

\begin{figure}
\centering
\subfigure[Near-critical Trajectories in the $Z$--$X$ plane.]{\label{fig:transition X}
\includegraphics[scale=0.28 ,angle=90]{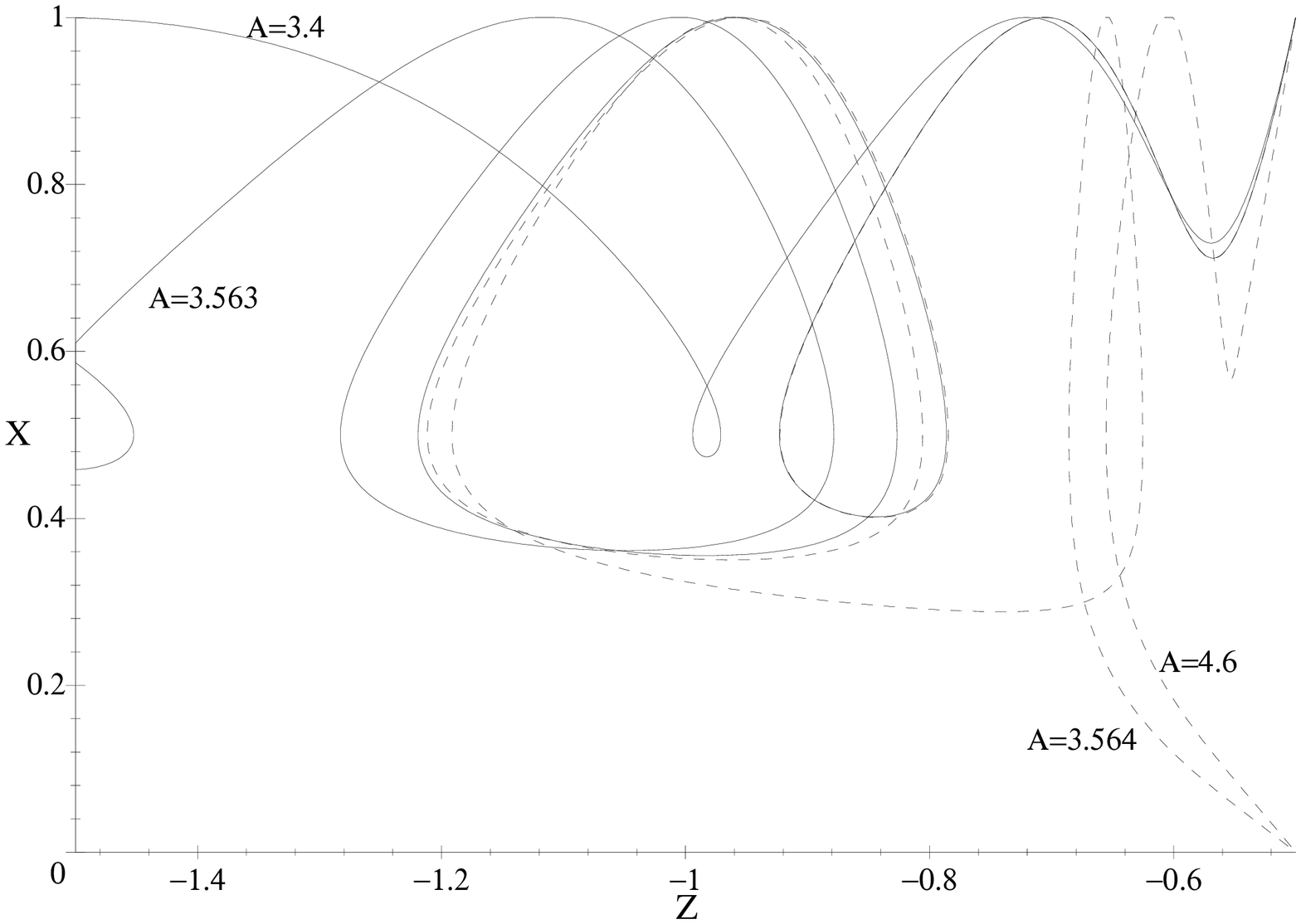}}
\hspace{0.3cm}
\subfigure[Near-critical Trajectories in the $Z$--$Y$ plane.]{\label{fig:transition Y}
\includegraphics[scale=0.28 ,angle=90]{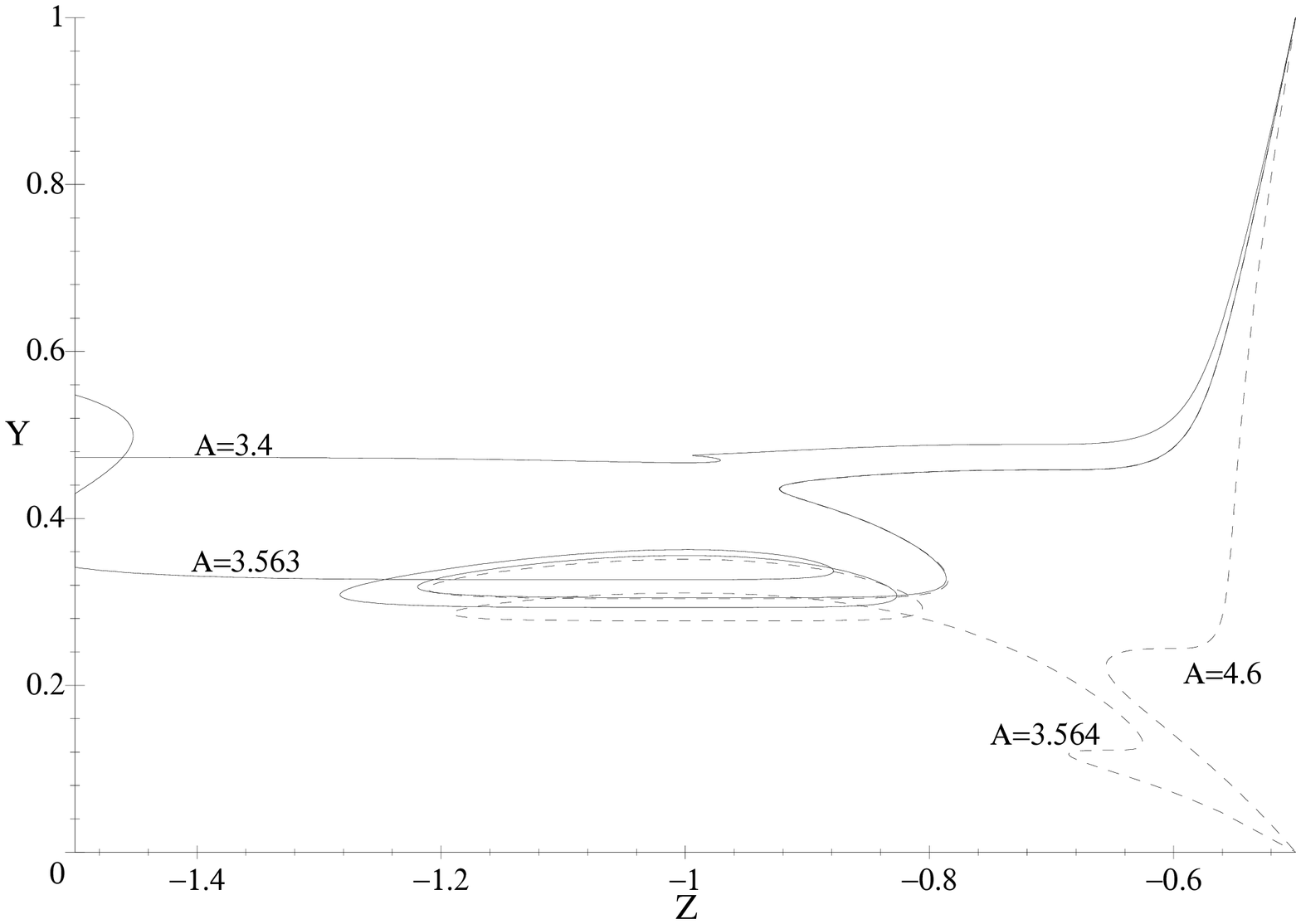}}
\label{fig:transition}
\caption{Near-critical Wyman trajectories; the sub-critical trajectories ($A=3.4$ and $A=3.563$) are shown as solid lines while the super-critical trajectories $A=3.564$ and $A=4.6$ are shown as dashed lines.}
\label{fig:near critical}
\end{figure}

As we see from Figure~\ref{fig:generic}, there is a qualitative transition at $A=A_1\approx 3.34$ where trajectories pass arbitrarily close to the point $\bigl(-1,\tfrac{1}{2}\bigr)$.
Whereas trajectories with $A<A_1$ display no tendency to collapse, those with $A>A_1$ undergo at least one period in which the gravitational attraction overcomes the tendency of the potential~\eqref{eq:NGT potential} to disperse the scalar field.
Trajectories with $A<A_1$ will never cross $X=\tfrac{1}{2}$, those with $A_1<A<A_c\approx 3.563$ will cross $X=\tfrac{1}{2}$ an even number of times, eventually propagating to $Z\rightarrow\infty$, and those with $A>A_c$ will cross $X=\tfrac{1}{2}$ an odd number of times, eventually ending at $\bigl(-\tfrac{1}{2},0\bigr)$.
Near-critical solutions ($A\approx A_c$) shown in Figure~\ref{fig:near critical} display an approximate periodicity, resulting from alternating periods in which either the gravitational attraction or the repulsive potential dominates.

\section{Discussion}
\vspace*{-0.5pt}
\noindent
Understanding these results is rather straightforward. 
In addition to the presence of the self--interaction term, the repulsive nature of~\eqref{eq:NGT potential} is acting quite strongly to disperse the field.
This is shown by the increase in the amount of energy required to produce a black hole over that of the E--KG system; $A_c=1$ for the E--KG system and $A_c\approx 3.563$ for the Wyman equations. 
We can also understand the oscillatory behaviour as the interplay of the attractive gravitational force and the repulsive potential.

Trajectories corresponding to very small initial energies ($A\approx 0$) are very close to their E--KG relatives; the field does not get large enough to feel an appreciable repulsive potential.
As we consider increasing initial energy (increasing $A$) the potential quickly manifests itself by `flattening out' the trajectories in Figure~\ref{fig:all Y} over those in Figure~\ref{fig:GR trajectories}; trajectories that were super-critical for the E--KG system have been repulsed by the Potential~\eqref{eq:NGT potential}.
Notice that the tendency of the potential to disperse the field also has the effect of decreasing the repulsive potential, so at some point the attractive gravitational potential will take precedence and the system will have a tendency to re-collapse.
For trajectories with initial momenta that is well above the critical value, enough energy is directed towards $r=0$ `fast enough' that a horizon forms before the repulsive potential has an opportunity to take effect (as for the E--KG system these are infinitely massive black holes, and the entire spacetime becomes trapped\cite{Brady:1994}).
Close to the critical trajectories, the system oscillates between attractive and repulsive phases many times before `choosing' its fate.
In a more general numerical simulation,\cite{Clayton+Demopoulos+Legare:1997} this behaviour should be straightforward to identify.

So far we have not discussed the ambiguity in the identification of the spacetime metric in UFT,\cite{Tonnelat:1982} an issue that is far from clear given the existence of multiple light cones.\cite{Lichnerowicz:1955,Maurer-Tison:1959}
For this system, the ambiguity vanishes for the following reasons:
For purely spherically symmetric fields the light cone is determined solely in terms of the metric in the $t$--$r$ sector, and, as usual, we only need to know this metric up to an arbitrary  conformal factor.
The ability to reduce the system to an E--KG system tells us that the (radial) causal structure determined from the $t$--$r$ components of the fundamental tensor therefore has physical meaning.
(It is also not difficult to show that the metrics determined by Maurer--Tison\cite{Maurer-Tison:1959} are all conformally equivalent to a metric with the same $t$--$r$ components discussed here.)
This unambiguously provides the causal boundary for Wyman sector field propagation, and results from GR based on radial null geodesics (such as the existence of a horizon) are therefore relevant for this problem.

\section{Acknowledgements}
\vspace*{-0.5pt}
\noindent
The author would like to thank the Natural Sciences and Engineering Research Council of Canada for a Postdoctoral Fellowship, as well as L. Demopoulos and J. L\'{e}gar\'{e} for discussions relating to this work.

\end{document}